# DUV (257 nm) LIDT of Anti-Reflective Coatings

Martin Mydlář[1], Hana Turčičová[1], Priyadarshani Narayanasamy[1*], Mihai-George Mureşan[1], Ondřej Novák[1], Jan Vanda[1], Tomas Tolenis[2] Jan Brajer[1], and Tomáš Mocek[1]

[1]*HiLASE Centre, Institute of Physics of the Czech Academy of Sciences, Dolní Břežany, Czechia*

[2]*ELI Beamlines Facility, Extreme Light Infrastructure ERIC, Dolní Břežany, Czechia*

* Corresponding Author: priya@hilase.cz

**Abstract**

Laser-induced damage threshold in the deep ultraviolet (DUV) was used to perform tests on AR coated windows intended for industrial processes. The 257 nm wavelength was generated from Yb:YAG fundamental (1030 nm) and the long-term LIDT stability (up to 100000 shots) was selected as representative for minimal industrial qualification. Selected setup for the fourth harmonics generation consists of two stages, generation of second harmonic and subsequently the fourth harmonic with 13.4% conversation efficiency. In both 1-on-1 and S-on-1 experiments, the optics coated with $Al_2O_3/SiO_2$ by ion beam sputtering technique demonstrated the highest LIDT value of 0.73 and 0.5 J/cm$^2$ respectively. Various damage morphologies, color centers, ballistic damage, delamination, and blistering were observed across the samples with respect to the 1-on-1 and S-on-1.

## 1. Introduction

Optical components in the deep ultraviolet (DUV) laser systems fail abruptly when the fluence exceeds the limiting threshold of antireflection (AR) coating. Limiting threshold defines the fluence at which the optics can be operated safely and in the present case it is for DUV 257 nm. DUV has an inevitable need in semiconductor lithography, spectroscopy, micromachining and biomedical application [1,2]. This wavelength has high photon energy, smaller achievable focal point and large absorption cross section for many materials. These applications highly rely on this wavelength [3], so the rise in the laser induced damage threshold (LIDT) directly influences the performance in this application. For transmitting high intensity through these windows require suppression of Fresnel reflections which is achieved by multilayer AR coatings, because these reflections can cause loss in power, create ghost images and damage in the optical component due to backward propagation [4]. But the durability of these coatings remains the prime factor in limiting the lifetime and reliability of DUV optics.

At DUV wavelength, the photon energy is high enough to trigger multiphoton absorption and activate the defect. This tradeoff is very intense, and the damage further intensifies when windows are subjected to picosecond pulses [5,6]. Under repeated irradiation, the energy gets hugely incubated, leading to failure of the substrate well below the single-shot threshold. There are considerable publications and results in literature in understanding the wavelengths in the IR [7], visible [8] and UV [9] wavelengths but in the DUV wavelength, the data are sparse and very few studies were done especially in the picosecond regime. The LIDT mechanism of the nanosecond interactions fundamentally differs from the picosecond regime. In the ns the LIDT is dominated by heat diffusion, defect assisted absorption and lattice heating [10]. This partial dissipation results in high LIDT value in ns but in contrast the ps pulses are short and the heat dissipation is highly negligible, and the damage is initiated by electronic processes like multiphoton absorption and avalanche ionization. The optical material threshold on picosecond is significantly lower even when the nanosecond threshold is higher. Commonly, the manufacturers provide LIDT data at the nanosecond pulse following ISO 21254, which is standard for industrial UV lasers. The LIDT value extrapolated for picosecond from nanosecond cannot be reliable because of the defect induced electronic excitation in the shorter pulses. Many studies show that the scaling laws are inaccurate below the ~20 ps underestimating the true susceptibility of coatings and substrates for the ultrafast damage. Therefore, the LIDT value represented for nanosecond is generally an overestimation for the safe picosecond laser operation and does not represent the optical threshold and durability of optics [11,12]. Thus, direct picosecond LIDT characterization is necessary for evaluating the correct risk assessment, coating durability, and long-term operation of picosecond DUV laser systems.

Direct laser of 257 nm wavelength is not possible in solid-state laser because most host crystals, coatings and optics used in the laser system will suffer from huge absorption and damage, and there is also no efficient way for transitions in this regime. Where the excimer lasers cannot provide the 257 nm wavelength. The other option is to use the Yb: based solid state lasers operating at 1030 nm to generate fourth harmonic generation of high-power picosecond laser. Albeit, this nonlinear conversion approach provides DUV pulses with low frequency efficiency, but the power and beam quality is sufficient for the above-mentioned application. For this generation, high quality and more delicate optics are required rather than standard 266 nm. In the present work, we used 257 nm wavelength generated from the powerful HiLASE Perla B 1030 nm picosecond laser system [13] to address the research gap by testing LIDT on AR coated windows.

Unlike previous DUV fatigue studies focused on individual coating materials, this work comparatively evaluates finished industrial multilayer AR windows under 257-nm, ~1.2 ps irradiation, combining single-pulse LIDT, fatigue evolution to 10^5 pulses, and damage-morphology analysis.

## 2. Materials and Methods

The laser source employed in this experimental work was the Perla B laser system, which operates as a diode-pumped solid-state laser (DPSSL) based on an Yb:YAG thin-disk configuration. This advanced system was developed in HiLASE, Czech Republic, and can deliver pulses at a fundamental wavelength of 1030 nm. The repetition rate of the system was set to 1 kHz, and after compression, the pulse duration reached 1.2 ps. The output beam exhibited excellent quality with an $M^2$ value of less than 1.2, while the pulse energy could be tuned up to a maximum of 10 mJ [13, 14]. The generated laser beam was guided through a laser beam distribution system (LBDS) [15] that directed the output toward the laser-induced damage threshold (LIDT) station [16]. Within this system, the output of Perla B was remotely managed using a pulse picker, while the energy of the delivered pulses was regulated through an attenuator. The attenuator consisted of a half-wave plate combined with a thin-film polarizer, which enabled fine adjustment of fundamental laser intensity before reaching higher harmonic setup.

The power dependence and spatial beam profiles of both the fundamental beam at 1030 nm and the frequency-doubled beam at 515 nm were investigated. In addition, fourth-harmonic generation (FHG) at 257 nm was achieved using BBO nonlinear crystal [17]. The DUV beam at 257 nm was then directed into an automated LIDT station, allowing tests to be performed in compliance with ISO 21254 standards [18]. A schematic of the FHG process was prepared to illustrate the experimental configuration. Inside the LIDT chamber, the ultraviolet laser beam was focused using a precision lens, and the damage threshold of anti-reflective (AR) coated windows under DUV irradiation was systematically measured. The schematic diagram of the harmonic generation and LIDT testing is provided in Figure 1.

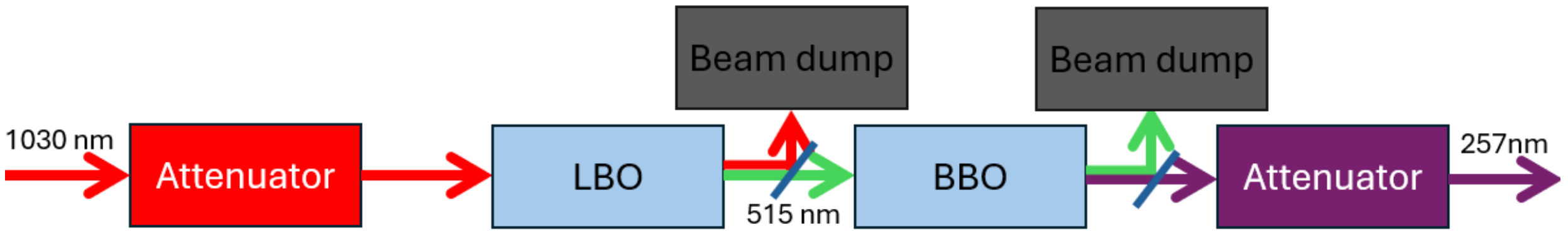

*Figure 1. Scheme of setup for FHG from 1030 nm to 257 nm. The power of output beam from Perla B system is controlled via an attenuator consisting of a half-wave plate and thin film polarizer. After attenuation, a second harmonic in LBO crystal is generated, resulting in a combination of 1030 nm and 515 nm. Residual fundamental 1030 nm is separated using a dichroic mirror and standalone wavelength 515 nm is up converted in BBO to 257 nm. To separate mix of 515 nm and 257 nm is used a dichroic mirror again and output power of 257 nm beam is controlled via attenuator.*

Prior to testing, the characteristics of the laser beam inside the chamber were thoroughly evaluated. Parameters such as pulse energy, beam profile, and spatial stability were recorded to ensure reliability of the measurements. The LIDT experiments were performed in a clean-room environment with an ISO class 6 cleanliness level, ensuring minimal contamination and high-quality testing conditions. The testing station itself was equipped with a precise positioning motorized sample mount and damage detection system. Each optical sample was secured on a motorized two-axis translational stage, which allows accurate movement of the specimen relative to the incident laser beam. An online damage-detection camera and white light source to see the scattering from the damage site is used to continuously monitor the laser-irradiated spot during exposure, allowing real-time observation of the initiation of damage. For further confirmation and detailed analysis of the damaged sites, a laser scanning microscope (OLS 5000, Olympus) was employed to examine the exposed areas after irradiation. The instrument used for the measurement of reflectance and transmittance of AR coated windows was UV-VIS-MWIR spectrophotometer (Photon RT UV-VIS-MWIR, Essent Optics). The measurements were done in the range of 200-400 nm with a sampling interval of 2 nm using a deuterium light source and the signal average integration is 10 ms. Sample thickness and the refractive index were set to 1 mm and 1.50 respectively.

AR-coated windows for 257 nm were tested using 2” optics at an angle of incidence (AOI) ≤5°. Samples were supplied in the framework of HiLASE 2025 LIDT Challenge [19] by different companies. Sample details are provided in Table 1. As shown in Figure 2, the generated 257 nm, with Gaussian beam diameter of 5.1 mm, in p-pol was focused by plano-convex $CaF_2$ lens (f = 1000 mm) to beam major diameter 306 µm and minor dimeter 275 µm at the $1/e^2$ intensity level. The fluence value for each attenuator position is calculated and recorded from the beam width and pulse energy for the LIDT experiment. Beam profile and energy were diagnosed using a CMOS camera (UI-5280CP, IDS) and power meter (UP25N-100H-H9-D0, Gentec). The online detection camera is focused on the surface of the coating to check the damage and prevent the catastrophic damage/contamination by debris on the optics. Post-mortem inspection was additionally used to identify less visible damaged spots (e.g. discoloration) by a laser scanning microscope. In the 2" optics, 8% of the clear aperture was used for the LIDT testing, which represented up to 367 testing sites. In the LIDT test, the ISO 21254 standard is followed for the 1-on-1 and S-on-1 (S = $10^5$ pulses) testing methodologies. The uncertainty budget included the uncertainty in beam diameter, estimated as the standard deviation of 20 independent measurements, and the uncertainty associated with pulse energy calibration. The energy calibration curve was obtained by varying the attenuator setting in 5% increments. A similar approach was previously used also in the case of HiLASE 2024 LIDT Challenge [20].

**Table 1**: List of tested samples together with the materials used for their fabrication and the applied coating techniques.

| Sample indentifier | Substrate material | Sample thickness [mm] | Low *n* coating material | High *n* coating material | Coating technology |
|---|---|---|---|---|---|
| S1 | Heraeus Suprasil 3002 | 3.08 | $SiO_2$ | $Al_2O_3$ | IBS |
| S2 | UVFS | 3.06 | $SiO_2$ | $Al_2O_3$ | IBS |
| S3 | UVFS | 6.33 | $SiO_2$ | $HfO_2$ | IBS |
| S4 | C7980 | 6.04 | $LaF_3$ | $MgF_2$ | e-Beam |
| S5 | UVFS | 5.02 | $SiO_2$ | $Al_2O_3$ | IBS |
| S6 | Fused silica | 6.51 | $SiO_2$ | $HfO_2$ | PIAD |
| S7 | Fused silica | 9.45 | Fluorides | Fluorides | e-Beam |

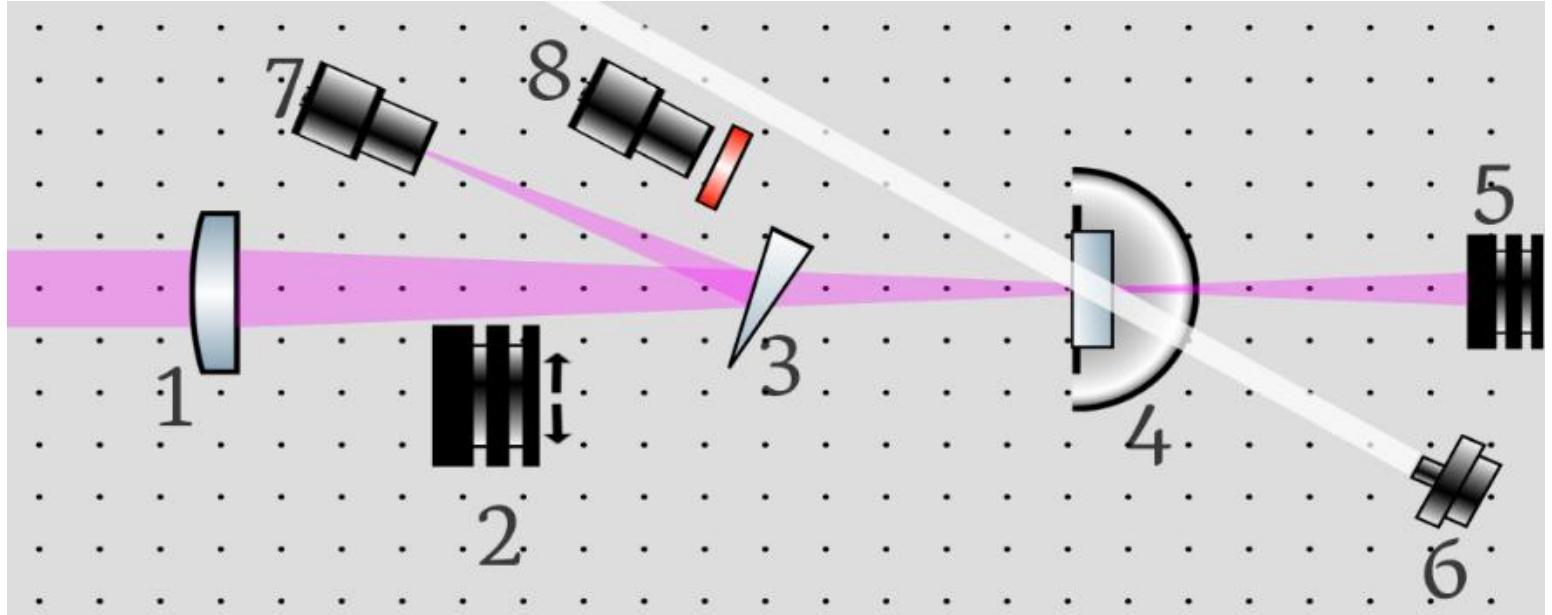

*Figure 2. LIDT testing setup scheme. 1 – focusing lens (f = 1000 mm), 2 – power meter, 3 – wedge, 4 – sample holder on motorized two-axis translational stage, 5 – beam dump, 6 – white light source, 7 – beam profiler, 8 – inline camera with low-pass filter*

## 3. Results and discussion

### 3.1 Generation of DUV

The frequency-conversion characteristics of the system were evaluated using a 1030 nm, 1.2 ps, 1 kHz, with LBO and BBO crystals employed for SHG and FHG, respectively. As shown in Figure 2, the SHG output from the LBO crystals (2 × 1.8 mm) exhibits linear increase with pump power, reaching approximately 4.5 W at 8 W of fundamental input. The corresponding SHG efficiency displays a rapid rise at low powers, followed by a gradual saturation trend beyond 6 W, achieving a maximum conversion efficiency of ~53.6%. This behavior is consistent with partial pump depletion and the onset of thermal or walk-off effects at higher intensities, while the overall trend reflects the favorable phase-matching bandwidth and low birefringent walk-off inherent to LBO [17].

For the FHG stage, the BBO crystal (0.75 mm) generates up to ~0.95 W of 4H output at the same 8 W input. The efficiency curve increases steadily to a peak of ~13.4%, after which a slight roll-off is observed. This saturation is attributed to the cumulative phase-mismatch in the cascaded 1030 → 515 nm → 257 nm process, along with the stronger angular sensitivity and larger walk-off associated with BBO. The linear regime at low powers suggests effective temporal and spatial overlap between the SHG and fundamental beams, whereas the reduced slope beyond the marked region indicates limited conversion at higher fluence.

Overall, the system demonstrates efficient and stable UV generation, with SHG and FHG efficiencies of approximately 53,6% and 13,4%, respectively, under identical pumping conditions. This corresponds to a net 1H→4H conversion of ~13,4%, or equivalently, ~25% of the generated 2H photons being further converted to 4H. These results confirm the suitability of the LBO–BBO configuration for high-power harmonic generation applications requiring short-pulse, high-repetition-rate ultraviolet output.

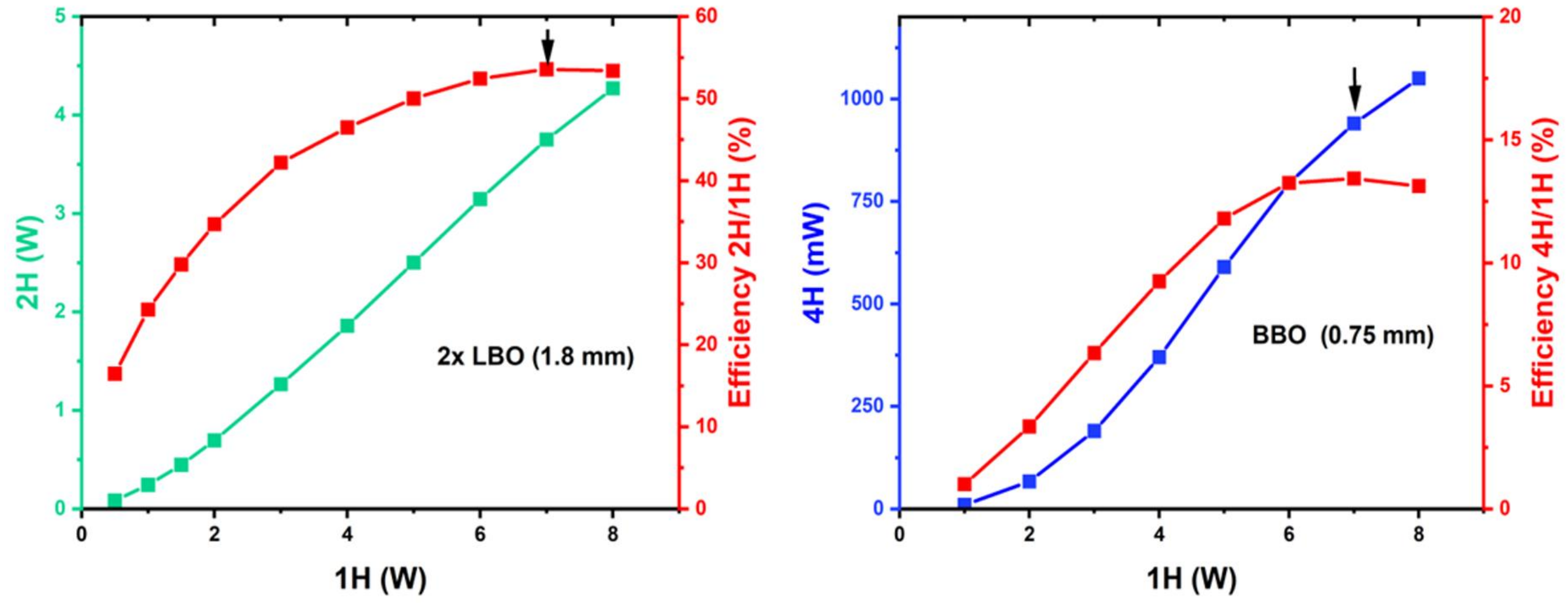

*Figure 3. Second-harmonic generation and conversion efficiency of fundamental beam from Yb:YAG thin-disk laser system using type I LBO crystal; (b) fourth-harmonic generation and conversion efficiency from BBO crystal.*

### 3.2 DUV Laser induced damage testing

In the LIDT, the samples were initially tested by 1-on-1 to determine the baseline fluence and save the sample from destruction and continued with S-on-1 method. 1-on-1 test means testing of each test side with single shot and S-on-1 determines multiple pulses at each site. Before irradiating multiple pulses on the sample, it is important to determine the maximum fluence at which the coatings can withstand 1-on-1 test. The 1-on-1 test was used to determine the single-pulse damage behavior and to define an appropriate fluence range for subsequent S-on-1 measurements. S-on-1 testing was then used to quantify the dependence of damage resistance on accumulated pulse number. For the S-on-1 (S=1, 2, 5, 10, 20, 50, 100, 200, 500, 1000, 10000, 100000) test, the number of pulses is gradually increased to have statistical data of damage threshold.

For each spot, the energies were set specifically with a fluence difference of 0.12 J/cm$^2$ for a systematic study. For statistical purposes, four sites are tested with the same condition, i.e., fluence and number of pulses. The sample's characteristic measurements are provided in the Table. 2 and the characteristic damage curve of the tested samples are presented below in Figure 4. For each combination of fluence and pulse number, four independent test sites were irradiated, and the damage probability was determined as the ratio of damaged sites to the total number of irradiated sites. The fluence was varied in increments of 0.12 J/cm$^2$. For each pulse class, the damage probability as a function of fluence was evaluated by linear regression within the damage-transition region. The LIDT was defined as the fluence corresponding to the extrapolated zero-damage probability ($P = 0$), i.e., the fluence-axis intercept of the linear fit. The values presented in Figure 4a therefore represent the extrapolated 0% damage-probability threshold as a function of the number of incident pulses.

**Table 2.** Measured optical properties of AR-coated windows (reflectance, transmittance), the calculated optical losses, and their LIDT results performed at 257 nm.

| Sample identifier | R [%] | T [%] | L [%] | LIDT 1-on-1 (J/cm$^2$) | LIDT $10^5$-on-1 (J/cm$^2$) |
|---|---|---|---|---|---|
| S1 | 0.26 | 99.55 | 0.19 | 0.70 | 0.41 |
| S2 | 0.13 | 99.72 | 0.15 | 0.73 | 0.50 |
| S3 | 0.04 | 99.50 | 0.46 | 0.63 | 0.37 |
| S4 | 3.98 | 95.48 | 0.54 | 0.54 | Substrate failure |
| S5 | 0.19 | 96.84 | 2.97 | 0.54 | 0.05 |
| S6 | 0.29 | 95.22 | 4.49 | 0.29 | 0.20 |
| S7 | 4.11 | 95.70 | 0.19 | 0.68 | 0.35 |

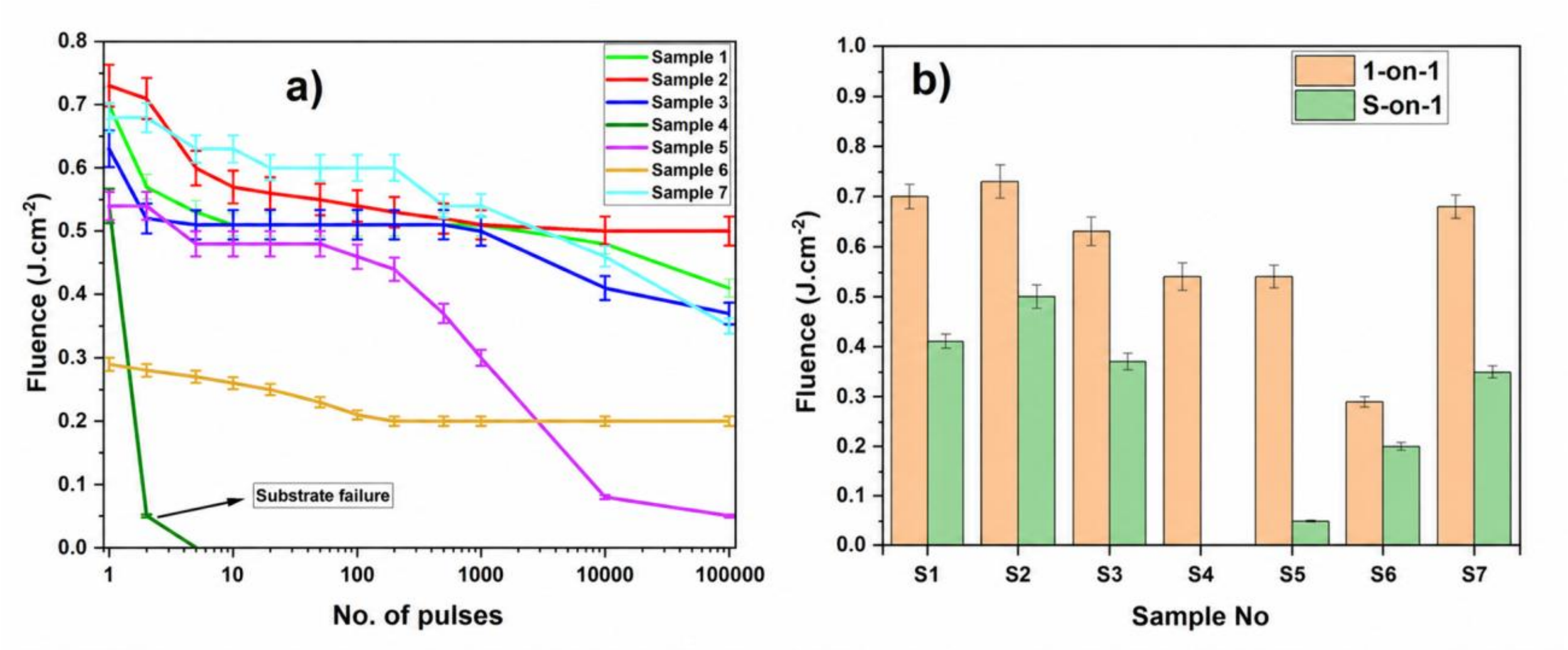

*Figure 4. LIDT results from the S-on-1 test: (a) characteristic damage curves extracted from the LIDT test; (b) comparison of 1-on-1 and S-on-1 LIDT value for hundred-thousand pulses on the coating.*

Owing to the short pulse duration and the relatively steep damage-probability transition observed under the present irradiation conditions, four independent sites were evaluated at each fluence level, providing a practical sampling density while allowing a large number of fluence and pulse-count combinations to be investigated over the available clear aperture.

Figure 4a demonstrates the evolution of LIDT value in regard to the number of pulses for each tested sample. In the 1-on-1 test, the LIDT value of the tested sample ranges from 0.29-0.73 J/cm$^2$ (Figure 4b). Sample S6 produced using a combination of $HfO_2/SiO_2$ shows the least LIDT value in 1-on-1 test (0.29 J/cm$^2$). Sample S2, coated with $Al_2O_3/SiO_2$, has the highest LIDT value in the same test configuration. On increasing the number of pulses, i.e., on the S-on-1 test, the LIDT value starts to decline and attains considerably lower value than in 1-on-1 test. Sample S2 and S6 showed a plateau which indicates a lack of degradation for a higher number of pulses. Sample S2 shows a high LIDT value (0.73 J/cm$^2$) in 1-on-1 test then slowly degrades and starts to be stable from 500 pulses to $10^5$ pulses (0.5 J/cm$^2$) to have a plateau of 500 pulses.

Sample S1 and S2 have the same coatings as $Al_2O_3/SiO_2$ by IBS technique. Though both the samples show the same LIDT value for the 1-on-1 test and have a similar trend in the curve until 1000 pulses, the coatings started to fail after 1000 pulses in sample S1. Sample S4, though it shows a considerably good LIDT value in 1-on-1 test, suffers from substrate failure [21].

The samples that were fabricated on the fused silica substrate (sample S1, S2, S3, S5 and S6) exhibit superior behavior for multiple pulses that indicate low defect density, minimal absorption, and high thermal conductivity of the substrate at 257 nm. However, the exception from the rule came from S4 (C7980 substrate), showed clear degradation for S-on-1 methodology, indicating defects or the precursor-driven fatigue laser damage.

Within the set of commercially supplied optics investigated here, the samples employing $Al_2O_3/SiO_2$ IBS coatings exhibited the highest measured damage resistance. Because coating designs and manufacturing parameters were not controlled between suppliers, the present measurements cannot isolate coating material or deposition method as the sole cause of this difference. Coatings based on $HfO_2$ (sample S3 and S6) demonstrated good 1-on-1 thresholds, but there is a general tendency towards noticeable fatigue at a higher number of pulses. This behavior aligns well with the known defect related to absorption in hafnia grown using PIAD process [22] like sample S6. It can be assumed that similar quality of the films for the IBS case sample S3. This may be due to the presence of suboxide centers that contribute to the gradual increase of absorption.

Deposition technique is crucial for thin film quality, which is another factor that influences LIDT performance. Samples S1, S2 and S3 are coated by IBS technique with low reflectance and high transmittance exhibited high LIDT value 0.3-0.5 $J/cm^2$. Although sample S1 has slightly higher reflectance than the other two samples, its superior transmittance and lower optical loss effectively compensate for the increase leading to overall LIDT performance. But the sample S5 coated by the same IBS techniques (R=0.2%, T= 96.8%) degraded in the S-on-1 test value to 0.05 $J/cm^2$. Even if the sample S2 and S5 are prepared by same material and technique, conditions like sample thickness, purity of the material, change in density will alter the reflectance and transmittance of the sample which influences its LIDT performance [23]. Thickness is the important parameter that directly influence the LIDT of the material where the thinner samples have higher LIDT than the thicker coating. Because the thicker sample is prone to higher defects while the thinner sample has high resistance against damage [24]. The high optical loss in sample S5 (2.97%) also directly affects the damage threshold (0.05 $J/cm^2$) which is not a case in sample S2 with an optical loss of 0.15%. and damage threshold of 0.5 $J/cm^2$. Among the currently discussed samples, Sample S1 and S2 (3.08 and 3.06 mm) are the thinnest samples, which achieved high LIDT value.

Sample S2 coated by IBS technique with low reflectance and high transmittance of 0.13% and 99.72% respectively shows highest LIDT. In general, thin films coated by IBS technique are amorphous and have low defect concentration, especially for materials like $HfO_2$ and $Al_2O_3$ which can have low optical losses due to low absorption in the UV range. Low optical losses (including absorption) can be correlated to high laser-damage threshold [25]. On the other hand, samples S4 and S7, which exhibited lower transmittance and higher reflectance, showed lower LIDT values. This was particularly evident for sample S4, whose coating rapidly deteriorated after only a few pulses. These results indicate that higher transmittance is associated with a higher LIDT.

Sample S6, coated using plasma ion-assisted deposition (PIAD), exhibited a lower 1-on-1 LIDT of 0.29 $J/cm^2$ and showed nearly the same fatigue behavior as the IBS-coated samples. PIAD usually provides a balance density and good deposition rate, but the oxygen deficient sites may change the LIDT value at multi-pulse damage due to collective absorption [26]. Sample S6 has higher losses 4.49% respectively, where the laser energy mostly transfers into heat and this is the reason for the drop in the LIDT value.

In general, fluoride material shows better LIDT performance in the DUV region compared to the oxide materials because they have higher bandgap [27]. But in the present work, the manufacturer of sample S7, deposited using e-beam technique, does not provide the exact coating material, mentioning only fluoride material. Although the sample S4 is coated with fluoride materials ($LaF_3$ and $MgF_2$), and using the same deposition technique, the coating failed abruptly after 10 pulses. So, the discussion on sample S4 cannot be concluded from its LIDT value. The reflectance of S4 and S7 is higher than the nominal specification for AR coatings and also higher than the other samples listed in Table. 2. Apparently, the e-beam technique for deposition of fluorides, especially DUV ranges are challenging to control. This can alter the thickness of the layers and therefore shift the spectral characteristics of the sample.

Overall, samples S2 and S6 exhibit a plateau during evolution in the number of pulses, with the damage threshold remaining the same up to $10^5$ pulses, where further accumulation will not significantly reduce the damage threshold. But in the case of other tested samples, the LIDT value decreases with an increase in pulse numbers indicating the incubation process is still active. Therefore, the material may exhibit cumulative fatigue, causing the LIDT to decrease with prolonged pulse exposure [28].

## 4. **Morphology of laser damage**

Laser parameters like wavelength, pulse length, repetition rate, number of pulses, spot size and fluence can influence the initiation of damage and damage type [29]. When laser interacts with the surface of the coating beyond it's damage threshold, the damage occurs. This damage may propagate the damage to other optics because of the hotspot formation in the initial damaged optics. In the present work, any faint visible change on the surface of the irradiated spot using Normanski microscope is considered as damage because this may develop into a ballistic damage after few hundred pulses, which reduces the anti-reflection properties and distorts the transmitted laser beam. DUV irradiation has higher photon energy, and if the photon energy equals or exceeds the bandgap of coating material, it can break the bond (photochemical damage) i.e., intrinsic damage in the coatings. For the sub-bandgap laser energies, photons initiate defect mediated absorption leading to the damage. [30]

Figure. 5 shows the various laser damage morphology of the samples (1-7) encountered during 1-on-1 (above) and S-on-1 (below) respectively. In both 1-on-1 and S-on-1, the sample exhibited damages ranging from pinhole, color centers, delamination, and small to big crater formation [31]. Even the small variation on the irradiated spot is considered damage because neglecting this damage will alter the LIDT value during real time usage of the optics and leads to catastrophic damage on the coatings. So delicate care is required in observing the damage, especially in the DUV optics. Sample 1 on the 1-on-1 test shows a pin hole damage on the irradiated fluence of 0.74 J/cm$^2$. A small pinhole can expand to form a big crater because of the defect induced damage on the surface on higher irradiation. The sharp edge pit with a minimal melting on the surrounding is seen on fluence of 0.62 J/cm$^2$ on sample 1, S-on-1. For S-on-1 methodology, even after lowering the fluence compared to 1-on-1 tests, the increased number of pulses created highly prominent damage. We also observed similar behavior of damages in sample 2 but on higher fluence than in sample 1. 1-on-1 on sample 3 shows stress-induced damage, that is the reason for the micro-crack on the coating surface. The threshold value is slightly higher at S-on-1 at 0.52 J/cm$^2$, 5 shots lead to delamination [32]. The minimal structure is modified on the surface due to the incubation stage.

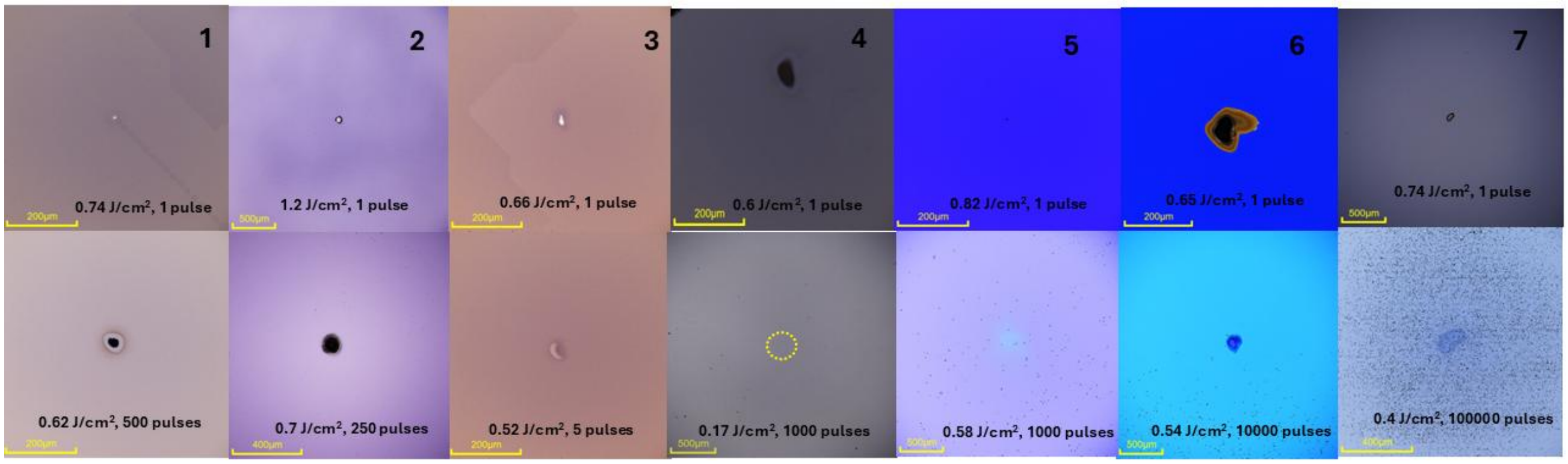

*Figure 5. Laser induced damage morphology of AR coated windows of Sample (1-7). The upper images represent 1-on-1 and lower images represent the S-on-1 test.*

In contrast, Sample 5, 1-on-1 shows the surface removal of the material with a halo around. On S-on-1 test, the damage is very faint with a small material removal with halo around, and the damage is highlighted for better view. Through the testing, this sample shows similar damage response even at a minimal fluence resulting in a failure of coatings and substrate. In sample 5, both 1-on-1 and S-on-1, the damage sites fluence value is very close to the damage threshold resulting in pinhole and color center [31]. The damage threshold in sample 6 is above the damage threshold resulting in the huge catastrophic crater at 1-on-1 test and smaller crater at 100000 pulses. Sample 7 1-on-1 damage threshold value is slightly higher than the tested value 0.74 J/cm$^2$ resulting in pinhole from the defect damage.

The repetition rate in the present case with the coating is 1ps which is in transition state where both the intrinsic and defect induced damages coexist. During DUV irradiation, the damages are mostly governed by the defect mediated electronic breakdown rather than the intrinsic breakdown contributes to higher fluence. Evolution of damages occurs in S-on-1 due to repeated incubation of energy, where repeated exposure enhances defect activation.

## 5. Conclusion

LIDT measurements of DUV AR-coated windows were performed at a wavelength of 257 nm, generated by frequency conversion of the 1030 nm output from the PERLA B laser system. At an average fundamental power of 8 W at 1030 nm, an average output power of 0.95 W at 257 nm was achieved, corresponding to an overall conversion efficiency of approximately 15% using LBO and BBO nonlinear crystals. The generated UV beam was focused onto the coated surface to a major beam diameter of 306 µm.

Seven DUV AR-coated window samples from different manufacturers were tested at 257 nm. Each irradiated site was continuously monitored using an online imaging system throughout the experiment. Among the tested samples, S2 and S1 exhibited the highest laser-induced damage thresholds in the S-on-1 test with $10^5$ pulses, reaching 0.50 J/cm$^2$ and 0.41 J/cm$^2$, respectively. Samples S5 and S6 demonstrated the most stable performance, showing consistent behavior during repeated irradiation. The sample consisting of an $Al_2O_3/SiO_2$ coating deposited by the ion beam sputtering (IBS) technique on a thin substrate also exhibited a high LIDT and stable performance during prolonged exposure.

The observed damage morphologies included color-center formation, coating delamination, pinhole defects, and crater formation, depending on the laser irradiation conditions and coating design. This work provides a comparative evaluation of the picosecond LIDT performance of commercially available DUV AR-coated windows at 257 nm, offering useful data for the selection of optical components for high-power DUV laser systems.

**Author Contributions: Martin Mydlář:** Writing – review & editing, Visualization, Methodology, Investigation, Formal analysis, Conceptualization. **Hana Turčičová:** Writing – review & editing, Visualization, Methodology, Investigation, Formal analysis, Conceptualization. **Priyadarshani Narayanasamy**: Writing – review & editing, Writing – original draft, Visualization, Methodology, Investigation, Formal analysis, Conceptualization. **Mihai-George Mureșan:** Formal analysis, Writing – review & editing, Validation. **Ondřej Novák:** Formal analysis, Writing – review & editing. **Jan Vanda**: Writing – review & editing, Validation, Supervision. **Tomas Tolenis:** Methodology, Formal analysis **Jan Brajer:** Supervision and **Tomáš Mocek:** Supervision

**Funding:** "This research was co-funded by the European Union and the state budget of the Czech Republic under the project LasApp CZ.02.01.01/00/22_008/0004573"

**Data Availability Statement:** The data will be made available upon legitimate request.

**Conflicts of Interest:** "The authors declare no conflicts of interest."